\documentclass[final]{siamltex}
\usepackage{graphics,amsmath,amsfonts,amssymb} 
\usepackage{epsfig,verbatim} 
\newtheorem{thm}{Theorem}[section]
\newtheorem{cor}[thm]{Corollary}
\newtheorem{ur:remark}{Remark}
\newenvironment{rem}{\begin{ur:remark}\rm}{\end{ur:remark}}

\def\sl{\mathrm{SL}}
\def\nc{\mathbb{C}}
\newcommand{\p}[1]{\left({#1}\right)}
\newcommand{\bm}[1]{\mbox{\boldmath ${#1}$}}

\title{Nonautonomous Hamiltonian Systems and Morales-Ramis Theory I. The Case \lowercase{$\ddot x=f(x,t)$}\thanks{Received by the editors xxxx; accepted for publication
(in revised form) by xxx; published electronically xxxx.}}

\author{Primitivo B. Acosta-Humanez\thanks{Departament de Matem\`atica Aplicada I, Universitat Polit\`ecnica de
Catalunya, Diagonal 647, E-08028 Barcelona - Spain
 ({\tt primitivo.acosta@upc.edu}).}}

\begin{document}

\maketitle

\begin{abstract}
In this paper we present an approach towards the comprehensive analysis of the non-integrability
of differential equations in the form $\ddot x=f(x,t)$ which is analogous to Hamiltonian systems with $1+1/2$ degree of freedom. In particular, we analyze the non-integrability of some important families of differential equations such as Painlev\'e II, Sitnikov and Hill-Schr\"odinger equation.
 We emphasize in Painlev\'e II, showing its non-integrability
through three different Hamiltonian systems, and also in Sitnikov
in which two different version including numerical results are
shown. The main tool to study the non-integrability of these kind
of Hamiltonian systems is Morales-Ramis theory. This paper is a
very slight improvement of the talk with the almost-same title
delivered by the author in SIAM Conference on Applications of
Dynamical Systems 2007.
\end{abstract}



\pagestyle{myheadings} \thispagestyle{plain} \markboth{Primitivo
Acosta-Hum\'anez}{Non Autonomous Hamiltonian Systems and
Morales-Ramis Theory I.}

\section{Introduction}
In this section we present the necessary theoretical background to understand the rest of the paper.
\subsection{Differential Galois Theory}\label{galois}

Our theoretical framework consists of a well-established crossroads of
Dynamical Systems theory, Algebraic Geometry and Differential
Algebra. See \cite{Morales} or \cite{SingerVanderput} for further
information and details. Given a linear differential system with
coefficients in $\mathbb{C}\!\left(t\right)$,
\begin{equation} \label{HODE}
\dot{\bm{z}}=A\left( t\right) \bm{z},
\end{equation}
a differential field $L\supset \mathbb{C}\!\left(t\right)$ exists,
unique up to $\mathbb{C}\!\left(t\right)$-isomorphism, which
contains all entries of a fundamental matrix
$\Psi=\left[\bm{\psi}_1,\dots,\bm{\psi}_n\right]$ of \eqref{HODE}.
Moreover, the group of differential automorphisms of this field
extension, called the \emph{differential Galois group} of
\eqref{HODE}, is an algebraic group $G$ acting over the
$\mathbb{C}$-vector space $\left\langle
\bm{\psi}_1,\dots,\bm{\psi}_n \right\rangle$ of solutions of
\eqref{HODE} and containing the monodromy group of \eqref{HODE}.

It is worth recalling that the integrability of a linear system
\eqref{HODE} is equivalent to the solvability of the identity
component $G^0$ of the differential Galois group $G$ of
\eqref{HODE} -- in other words, equivalent to the \emph{virtual
solvability} of $G$.

It is well established (e.g. \cite{mo1}) that any linear differential equation system
with coefficients in a differential field $K$
\begin{equation}\label{sisor1} \frac d{dt}\left(\begin{array}{c} \xi_1 \\
 \xi_2
\end{array}\right) =\left(\begin{array}{cc} a(t) & b(t)  \\ c(t) & d(t)
  \end{array}\right)\left(\begin{array}{c} \xi_1 \\
 \xi_2
\end{array}\right),
\end{equation}
by means of an elimination process, is equivalent to the second-order equation
\begin{equation}\label{sisor2}
\ddot \xi-\left(a(t)+d(t)+\frac {\dot b(t)} {b(t)}\right)\dot \xi -\left(\dot
a(t)+b(t)c(t)-a(t)d(t)-\frac{a(t)\dot b(t)}{b(t)}\right)\xi=0,
\end{equation}
where $\xi:=\xi_1$. Furthermore, any equation of the form $\ddot z-2p\dot z-qz=0$,
can be transformed, through the change of variables $z=ye^{{\int}p}$,
into $\ddot y=-ry$, $r$ satisfying the Riccati equation $\dot p=r+q+p^2$. This change is useful since it restricts
the study of the Galois group of $\ddot y=-ry$ to that of the algebraic subgroups of $\sl_2\p{\nc}$.

A natural question which now arises is to determine what happens
if the coefficients of the differential equation are not all
rational. A new method was developed in \cite{acbl}, in order to
transform a linear differential equation of the form $\ddot
x=r(t)x$, with transcendental or algebraic non-rational
coefficients, into its algebraic form -- that is, into a
differential equation with rational coefficients. This is called
the \emph{algebrization method} and is based on the concept of
\emph{Hamiltonian change of variables} \cite{acbl}. Such a change
is derived from the solution of a one-degree-of-freedom classical
Hamiltonian.
\\

\begin{definition}[Hamiltonian change of variables]\label{defi2} A change of variables $\tau=\tau\p{t}$ is called \textbf{Hamiltonian} if
$\p{\tau(t), \dot\tau\p{t}}$ is a solution curve of the autonomous
Hamiltonian system $X_H$ with Hamiltonian function
$$H=H(\tau,p)={\frac{p^2}{2}}+\widehat{V}(\tau),\textit{ for some } \widehat{V}\in \mathbb
C(\tau).$$
\end{definition}

\begin{thm}[Acosta-Bl\'azquez algebrization method \cite{acbl}]\label{pr2}
Equation $\ddot x=r(t)x$ is algebrizable by means of a Hamiltonian change of variables
$\tau=\tau(t)$ if, and only if, there exist $f,\alpha$ such that
${\frac{d}{d\tau}}\left(\ln \alpha\right),{\frac{f}{\alpha}}\in \mathbb{C}(\tau),$
where $$f(\tau(t))=r(t),\quad \alpha(\tau)=2(H-\widehat{V}(\tau))=(\dot\tau)^2.$$
Furthermore, the algebraic form of
$\ddot x=r(t)x$ is
\begin{equation}\label{equ4}
\frac{d^2 x }{d\tau^2}+\left(\frac{1}{2}\frac{d}{d\tau}\ln \alpha\right)\frac{dx}{d\tau}-\left(\frac{f}{\alpha}\right)x=0.  \quad\square
\end{equation}
\end{thm}

The next intended step, once a differential equation has been
algebrized, is studying its Galois group and, as a causal
consequence, its integrability. Concerning the latter, and in
virtue of the invariance of the identity component of the Galois
group by finite branched coverings of the independent variable
(Morales-Ruiz and Ramis, \cite[Theorem 5]{moralesramis}), it was
proven in \cite[Proposition 1]{acbl} that the identity component
of \emph{the Galois group is preserved in the algebrization
mechanism}.

The final step is analyzing the behavior of $t=\infty$ (or $\tau=\infty$) by studying the behavior of $\eta=0$
through the change of variables $\eta=1/t$ (or $\eta=1/\tau$) in the transformed differential equation, i.e.
$t=\infty$ (or $\tau=\infty$) is an ordinary point (resp. a regular singular point, an irregular singular point) of
the original differential equation if, and only if, $\eta=0$ is one such point for the transformed differential equation.

\subsection{Morales-Ramis Theory}
\emph{Everything is considered in the complex analytical setting from now on.} The heuristics of
the titular theory rest on the following general principle: if we assume system
\begin{equation}\label{DS} \bm{\dot{z}}=X\left(\bm{z}\right)
\end{equation}
``integrable" in some reasonable sense, then the corresponding
variational equations along any integral curve
$\Gamma=\left\{\widehat{\bm{z}}\left(t\right):t\in I\right\}$ of
\eqref{DS}, defined in the usual manner
\begin{equation}\label{vegamma}\tag{$\mathrm{VE}_{\Gamma}$}
\dot{\bm{\xi}}=X'\left(\widehat{\bm{z}}\left(t\right)\right)\bm{\xi},
\end{equation}
must be also integrable -- in the Galoisian sense of the last
paragraph in \ref{galois}. We assume $\Gamma$, a Riemann surface,
may be locally parametrized in a disc $I$ of the complex plane; we
may now complete $\Gamma$ to a new Riemann surface
$\overline{\Gamma}$, as detailed in \cite[\S 2.1]{moralesramis}
(see also \cite[\S 2.3]{Morales}), by adding equilibrium points,
singularities of the vector field and possible points at infinity.

The aforementioned ``reasonable'' sense in which to define
integrability if system \eqref{DS} is \emph{Hamiltonian} is obviously the one
given by the Liouville-Arnold Theorem, and thus the above general
principle does have an implementation:
\\

\begin{thm}[J. Morales-Ruiz \& J.-P. Ramis, 2001] \label{moralesramis}
Let $H$ be an $n$-degree-of-freedom Hamiltonian having $n$
independent rational or meromorphic first integrals in pairwise
involution, defined on a neighborhood of an integral curve
$\overline{\Gamma}$. Then, the identity component
$\mathrm{Gal}\left(\mathrm{VE}_{\overline{\Gamma}}\right)^0$ is an
abelian group (i.e.
$\mathrm{Gal}\left(\mathrm{VE}_{\overline{\Gamma}}\right)$ is
\emph{virtually abelian}).
\end{thm}
\\

The disjunctive between \emph{meromorphic} and \emph{rational}
Hamiltonian integrability in Theorem \ref{moralesramis} is related
to the status of $t=\infty$ as a singularity for the normal
variational equations. More specifically, and besides the
non-abelian character of the identity component of the Galois
group, in order to obtain Galoisian obstructions to the
\emph{meromorphic} integrability of $H$ the point at infinity must
be a regular singular point of \eqref{vegamma}. On the other hand,
for there to be an obstruction to complete sets of \emph{rational}
first integrals, $t=\infty$ must be a irregular singular point.
\\

See \cite[Corollary 8]{moralesramis} or \cite[Theorem
4.1]{Morales} for a precise statement and a proof.

\subsection{Non Autonomous Hamiltonian Systems}

\emph{Non-autonomous Hamiltonian systems} on symplectic manifolds
have long been the subject of study, and appear in a most natural
way in Classical Mechanics and Control Theory, e.g.
\cite{AbrahamMarsden}, \cite{Arnold}, \cite{Kuwabara},
\cite{LichtenbergLieberman}, \cite{Lanczos}, \cite{stiefscheif},
\cite{Stump}, \cite{Thirring}.

We consider non-autonomous Hamiltonian systems of the form
\begin{equation}\label{NAHS}
H=H(q_1,p_1,t)=\frac{p_1^2}{2}+V(q_1,p_1,t),
\end{equation}
$H$ is a non-autonomous Hamiltonian system with $1+1/2$ degree of
freedom. It is well-known (e.g. \cite{stiefscheif}) that
\eqref{NAHS} can be included as a subsystem of the Hamiltonian
system with two degrees of freedom given by
\begin{equation}\label{hs2}
\widehat{H}=\widehat{H}(q_1,q_2,p_1,p_2)=\frac{p_1^2}{ 2}+V(q_1,p_1,q_2)+p_2,
\end{equation}
where $q_2$ and $p_2$ are conjugate variables, i.e. $p_2=-H+k$,
where $k$ is constant, and $q_2=t$. Furthermore, $p_2$ is easily
seen to be the offset or counterbalancing energy of the system
(\cite{stiefscheif}, \cite{Struck}).

Also worth mentioning are some recent results on canonical
transformations in the extended phase space \cite{Struck}, \cite{StruckRied1},
\cite{StruckRied2}, \cite{Tsiganov}, as well as on
definitions and consequences of ``integrability'' under such
circumstances or generalizations thereof, even for non-Hamiltonian
systems (\cite{BD}, \cite{BDL}, \cite{GMS}, \cite{Kaushal}) which
we will not delve into further at this point.

\section{Main results}\label{mainresults}

Consider the differential equation
\begin{equation}\label{eq1}
\ddot x = f(x,t),
\end{equation} with particular solution $x=x(t)$. We will henceforth order our choice of positions as
$q_1=x$ and $q_2=t$, thus yielding a Hamiltonian system given by
$$ H = {\frac{p_1^2}2}-
F\left(q_1,q_2\right), \quad F_{q_1}\left(q_1,q_2\right)= \frac{\partial
F\left(q_1,q_2\right)}{
\partial q_1 }=f(q_1,q_2).$$ Equation \eqref{eq1} is obviously equivalent
to Hamilton's equations for $H$,
$$
\dot q_1=p_1=H_{p_1}\qquad \dot p_1=-H_{q_1}= f(q_1,q_2);
$$
this non-autonomous Hamiltonian system is included as a subsystem
of $X_{\widehat{H}}$ linked to $\widehat{H}:=H+p_2$, such as in
equation \eqref{hs2}. Assuming $x(t)=q_1(t)$ to be a solution of
\eqref{eq1} and $q_2(t)=t$, we obtain an integral curve
$\Gamma=\left\{\bm{z}\left(t\right)\right\}$ of $\widehat{H}$,
where
$$\bm{z}\left(t\right):=\left(q_1(t),q_2(t),p_1(t),p_2(t)\right)=\left(q_1(t),t,\dot{q}_1(t),-H(t)\right).$$
We may now introduce our first main result:
\\

\begin{thm}\label{th1} Let $\Gamma$ be an integral curve of $X_{\widehat{H}}$ such as the one
introduced above. If $X_{\widehat{H}}$ is integrable by means of
rational or meromorphic first integrals, then the Galois group of
\begin{equation}\label{eq3}
\ddot \xi= \left(f_{q_1}(q_1,q_2)|_{\Gamma}\right)\xi,
\end{equation}
is virtually abelian.
\end{thm}
\\

\begin{proof}
The Hamiltonian field $X_{\widehat{H}}$ is given by $
X_{\widehat{H}}=\left(
  p_1,
 1, f(q_1,q_2),
  F_{q_2}(q_1,q_2)
\right)^T. $ The variational equation $\mathrm{VE}_{\Gamma}$ along
$\Gamma$ is
\begin{equation}\label{eq4}
\frac{d}{dt}\left(\begin{array}{c}
  \xi_1 \\
  \xi_2 \\
  \xi_3 \\
  \xi_4
\end{array}\right)=\left.
 \left(\begin{array}{cccc}
  0 & 0 & 1 & 0 \\
0 & 0 & 0 & 0 \\
  f_{q_1}(q_1,q_2) & f_{q_2}(q_1,q_2) & 0 & 0\\
    F_{q_1q_2}(q_1,q_2) & F_{q_2q_2}(q_1,q_2) & 0 & 0
\end{array}\right)\right|_{\Gamma}\left(\begin{array}{c}
  \xi_1 \\
  \xi_2 \\
  \xi_3 \\
  \xi_4
\end{array}\right),
\end{equation}
More precisely, equation (\ref{eq4}) is
\begin{equation}\label{eq5}
\left\{
\begin{array}{lll}
  \dot \xi_1 & = & \xi_3, \\
  \dot \xi_2 & = & 0, \\
  \dot \xi_3 & = & \left(f_{q_1}(q_1,q_2)|_{\Gamma}\right)\xi_1+\left(f_{q_2}(q_1,q_2)|_{\Gamma}\right)\xi_2, \\
  \dot \xi_4 & = & \left(F_{q_1q_2}(q_1,q_2)|_{\Gamma}\right)\xi_1+\left(F_{q_2q_2}(q_1,q_2)|_{\Gamma}\right)\xi_2.
\end{array}\right.
\end{equation}
Hence $\xi_2=k$, where $k$ is constant. Assuming $k=0$, the
\emph{normal variational equations} ($\mathrm{NVE}_{\Gamma}$, see
\cite[\S 1]{BoucherWeil}, \cite[\S 4.3]{moralesramis}, \cite[\S
4.1.3]{Morales}) for $\widehat{H}$ are given by
\begin{equation}\label{eq6}
\frac{d}{dt}\left(\begin{array}{c}
 \xi_1 \\ \xi_3
\end{array}\right)=\left(\begin{array}{cc}
  0 & 1 \\
  f_{q_1}(q_1,q_2)|_{\Gamma} & 0
\end{array}\right)\left(\begin{array}{c}
 \xi_1 \\ \xi_3
\end{array}\right),
\end{equation}
and solving \eqref{eq6} we can obtain $\xi_4$. System \eqref{eq6}
is equivalent to equation \eqref{eq3}, where $\xi=\xi_1$. In virtue of
\cite[Proposition 4.2]{Morales}, the virtual commutativity of
$\mathrm{Gal}\p{\mathrm{VE}_{\Gamma}}$ implies that of
$\mathrm{Gal}\p{\mathrm{NVE}_{\Gamma}}$; this, coupled with
Theorem \ref{moralesramis}, implies that if $X_{\widehat{H}}$ is rationally or meromorphically
integrable, then the Galois group of the equation \eqref{eq3}
is virtually abelian.
\end{proof}
\\

\begin{cor}\label{cor1} Suppose the Galois group of the differential equation
$\ddot \xi=k(t)\xi$ not virtually abelian. Define
$H:=\frac{p_1^2}{2}-k(q_2)\frac{q_1^2}{2}$ and
$\widehat{H}:=H+p_2.$ Then $X_{\widehat{H}}$ is not integrable by
means of meromorphic or rational first integrals. $\square$
\end{cor}
\\

\begin{rem} If the Galois group of the equation $\ddot \xi=k(t)\xi$ is
the Borel group $G=\mathbb{C}^*\ltimes\mathbb{C}$
 (hence connected, solvable and non-abelian),
$X_{\widehat{H}}$ is neither meromorphically nor rationally
integrable, although it is still possible to solve the equation --
as well as, ostensibly, the non-autonomous Hamiltonian system
$X_H$.
\end{rem}
\\

In anticipation of the following Corollary consider, for any
$g(x)$, $a(t)$ and $\alpha=\alpha_0$ given, the equation

\begin{equation}\label{eq7}
\ddot x = g_x(x)(g(x)+a(t))+\alpha,\quad \alpha\in\mathbb{C},
\end{equation}
having a certain known particular solution $q_1=q_1\p{t}$. Let $$H = \frac{p_1^2}{2}-
\frac{\left(g(q_1)+a(q_2)\right)^2}{2}-\alpha q_1, \quad q_2=t$$ be a Hamiltonian
linked to \eqref{eq7}, and
$\widehat{H}:=H+p_2$ its autonomous completion.
\begin{cor}\label{cor2} If $X_{\widehat{H}}$
is integrable through rational or meromorphic first integrals
then, along the integral curve
$\Gamma=\left\{\bm{z}(t)=\p{q_1(t),t, \dot{q}_1(t),-H(t)}\right\}$, the Galois group
of the equation
\begin{equation}\label{eq9}
\ddot \xi=\left.
\left(g^2_{q_1}(q_1)+g_{q_1q_1}(q_1)(g(q_1)+a(q_2))\right)\right|_{\Gamma}\xi,
\end{equation}
is virtually abelian. $\square$
\end{cor}
\\

Now, keeping the above hypotheses for $g(x)$, $a(t)$, $\alpha=\alpha_0$,
equation (\ref{eq7}) and $x=q_1\!\p{t}$, let us define the Hamiltonian
system
\begin{equation}\label{eq10}
\widehat{H}=H+p_2,\quad H=\frac{p_1^2}{2}-(g(q_1)+a(q_2))p_1-(\alpha+
a_{q_2}(q_2))q_1,\quad \alpha\in\mathbb{C}.\end{equation}
It only takes the following simple calculation to prove that Hamiltonian $H$ is indeed linked
to \eqref{eq7} in the manner expected, i.e. as introduced in the beginning of Section \ref{mainresults}.
We have
$$\begin{array}{lll}
\ddot x &=& g_x(x)(g(x)+a(t))+\alpha\\
&=& g_x(x)\dot x+ g_x(x)g(x)+g_x(x)a(t)+\alpha+\dot a(t)-g_x(x)\dot x-\dot a(t)\\
&=&g_x(x)y+\alpha+\dot a(t)-\frac{d}{dt}(g(x)+ a(t)),
\end{array}
$$
where $y=\dot x+ g(x)+a(t)$, and requiring $x$ and $y$ to be conjugate variables implies
the following system, equivalent to \eqref{eq7},
$$\begin{array}{lllll}
\dot x &=& y-\p{g(x)+a(t)}&=&H_y\\
\dot y&=&g_x(x)y+\alpha+\dot a(t)&=&-H_x.
\end{array}
$$
A straightforward parallel integration,
$$
H=\frac{y^2}{2}-\left(g(x)+a(t)\right)y+h_1(x,t) = h_2(y,t)-g(x)y-(\alpha+\dot a(t))x,
$$
along with the definition of
 $$h_1(x,t)=-(\alpha+\dot a(t))x,\quad h_2(y,t)=\frac{y^2}{2}-a(t)y,$$
as well as $\p{q_1,q_2,p_1}=\p{x,t,y}$,
yields the autonomous Hamiltonian system introduced in \eqref{eq10}.

\begin{thm}\label{th2} If $X_{\widehat{H}}$ is integrable by means of rational or meromorphic first integrals,
then, along $\Gamma=\left\{\bm{z}\p{t}=\p{q_1(t),t,p_1(t),-H(t)}\right\}$, the
Galois group of the equation
\begin{equation}\label{eq11}
\ddot \xi=\left.
\left(g^2_{q_1}(q_1)+p_1g_{q_1q_1}(q_1)-g_{q_2q_1}(q_1)\right)\right|_\Gamma\xi,
\end{equation}

is virtually abelian.
\end{thm}
\\

\begin{proof}
We may proceed as in the proof of Theorem \ref{th1}.
The Hamiltonian field $X_{\widehat{H}}$ is given by
$$X_{\widehat{H}}=\left(\begin{array}{c}
  p_1-(g(q_1)+a(q_2))\\
 1\\ g_{q_1}(q_1)p_1+(\alpha+a_{q_2}(q_2))\\
 a_{q_2}(q_2)p_1+a_{q_2q_2}(q_2)q_1\end{array}
\right). $$
The variational equation $\mathrm{VE}_{\Gamma}$ along $\Gamma=\left\{(q_1(t),t,p_1(t),-H(t))\right\}$ is
\begin{equation}\label{eq41}
 \bm{\dot\xi}=
 \left.
 \left(\begin{array}{cccc}
  -\frac{\partial g(q_1)}{\partial q_1} & -\frac{\partial a(q_2)}{\partial q_2} & 1 & 0 \\
0 & 0 & 0 & 0 \\
  p_1\frac{\partial^2 g(q_1)}{\partial q_1^2} & \frac{\partial^2 a(q_2)}{\partial q_2^2} & \frac{\partial g(q_1)}{\partial q_1} & 0\\
    \frac{\partial^2 a(q_2)}{\partial q_2^2} & \left(p_1\frac{\partial^2a(q_2)}{\partial q_2^2}+q_1\frac{\partial^3 a(q_2)}{\partial q_2^3}\right) & {\partial a(q_2)\over \partial q_2} & 0
\end{array}\right)
\right|_{\Gamma}\bm{\xi},
\end{equation}
where $\bm{\xi}=  \left(\xi_1,\xi_2,\xi_3,\xi_4\right)^T$. $\xi_2\equiv k\in\mathbb{C}$; in particular,
$k=0$ renders system \eqref{eq41} equal to the following:
\begin{equation}\label{eq51}
\left\{
\begin{array}{lll}
  \dot \xi_1 & = & \left.-g_{q_1}(q_1)\right|_{\Gamma}\xi_1+\xi_3, \\
  \dot \xi_2 & = & 0, \\
  \dot \xi_3 & = & \left.p_1g_{q_1q_1}(q_1)\right|_{\Gamma}\xi_1+\left.-g_{q_1}(q_1)\right|_{\Gamma}\xi_3, \\
  \dot \xi_4 & = & \left.a_{q_2q_2}(q_2)\right|_{\Gamma}\xi_1+\left.a_{q_2}(q_2)\right|_{\Gamma}\xi_3,
\end{array}\right.
\end{equation}
$\mathrm{NVE}_{\Gamma}$ corresponding to
\begin{equation}\label{eq61}
\frac{d}{dt}\left(\begin{array}{c}
 \xi_1 \\ \xi_3
\end{array}\right)=\left.\left(\begin{array}{cc}
 - g_{q_1}(q_1) & 1 \\
 p_1g_{q_1q_1}(q_1) & g_{q_1}(q_1)
\end{array}\right)\right|_{\Gamma} \left(\begin{array}{c}
 \xi_1 \\ \xi_3
\end{array}\right),
\end{equation}
and solving \eqref{eq61} we obtain $\xi_4$.
Using equations \eqref{sisor1} and \eqref{sisor2}, system \eqref{eq61}
is equivalent to equation \eqref{eq11}, where $\xi=\xi_1$. Again in virtue of \cite[Proposition 4.2]{Morales}
as in Theorem \ref{th1}, the integrability of Hamiltonian
$\widehat{H}$ by means of meromorphic or rational
first integrals implies the virtual commutativity, of the Galois group of equation \eqref{eq11}.
\end{proof}
\\

\begin{rem}
We observe that if $\ddot x=f(x,t)$, and if the same particular
solution $x=x(t)$ occurs in the statements of Theorem \ref{th1},
Corollary \ref{cor2} and Theorem \ref{th2}, then we obtain the
same ${\mathrm{NVE}_{\Gamma}}$ (equations (\ref{eq3}), (\ref{eq9})
and (\ref{eq11}) are equivalent), despite the fact that their
respective linked Hamiltonian systems have different expressions.
\end{rem}
\\

\section{Examples}
In this Section, and in application of Theorems \ref{th1} and \ref{th2} as well as of Corollaries \ref{cor1} and \ref{cor2},
we analyze the non-integrability of the Hamiltonian systems corresponding to the following differential equations:
\\

\begin{enumerate}
\item Hill-Schr\"odinger equation: $\ddot x=k(t)x$,
\item The Sitnikov problem: $\ddot x=-{(1-e\cos t)x \over (x^2+r^2(t))^{3 \over 2}},$ $r(t)={1-e\cos t \over 2}.$
\item Painlev\'e II equation: $\ddot x=2x^3+tx+\alpha,$
\item An algebraic toy model: $\ddot x=-{1\over 4x^3}-{t\over x^2}+\alpha,$
\end{enumerate}

In order to analyze normal variational equations, a standard
procedure is using \textsc{Maple}, and especially commands
\texttt{dsolve} and \texttt{kovacicsols}. Whenever the command
\texttt{kovacicsols} yields an output ``\texttt{[ ]}", it means
that the second-order linear differential equation being
considered has no Liouvillian solutions, and thus its Galois group
is virtually non-solvable. For equations of the form $\ddot y=ry$
with $r\in\mathbb{C}(x)$ the only virtually non-solvable group is
$\sl_2\p{\nc}$. In some cases, moreover, \texttt{dsolve} makes it
possible to obtain the solutions in terms of special functions
such as \emph{Airy functions}, \emph{Bessel functions} and
\textbf{hypergeometric functions}, among others
(\cite{Abramowitz}). There is a number of second-order linear
equations whose coefficients are not rational, and whose solutions
\textsc{Maple} cannot find by means of the commands
\texttt{dsolve} and \texttt{kovacicsols} alone; this problem, in
some cases, can be solved by the stated algebrization procedure.

\subsection{Hill-Schr\"odinger equation $\ddot x=k(t)x$} This example corresponds to Corollary \ref{cor1}.
For $k>0,\epsilon\gg0$, $\mathcal{P}_{n}$ a polynomial of degree
$n$ with $\mathcal{P}_{n}(0)=1$ and $k(t)$ given by
$$\begin{array}{ll}k(t)=k{\rm{e}}^{-\epsilon t},&k(t)=k\mathcal{P}_n(\epsilon t),\\
k(t)=k(1+\sinh(\epsilon t)),&k(t)=k(1+\sin(\epsilon t)),\\
k(t)=k(1+\cosh(\epsilon t)),&k(t)=k(1+\cos(\epsilon t)).
\end{array}
$$
$X_{\widehat{H}}$ is non-integrable by means of rational first integrals.

The integrability of equation $\ddot x=k(t)x$ for these examples has been deeply analyzed in \cite{acbl}.
\\

\subsection{The Sitnikov problem\label{NonApprox}}

The \emph{Sitnikov problem} is a symmetrically configured restricted three-body problem in which two primaries
with equal masses move in ellipses of eccentricity $e$ in a plane $\pi _1$, and an infinitesimal point mass moves
along the line $\pi _1^{\perp }$. See \cite{PhysRevA.14.2338}, \cite{HeviaRanada}, \cite{moralesramis}, \cite{Moser},
\cite{Whittaker}, \cite{Wodnar}, \cite{HagelTrenkler} for more details. The motion of the infinitesimal point mass is
given by the following differential equation
\begin{equation}
\ddot{z}+\frac z{\left( r^2\left( t\right) +z^2\right) ^{3/2}}=0,
\label{sitnikov}
\end{equation}
where $z=z\left( t\right) $ is the distance from the infinitesimal mass point to the plane of the primaries and
$r\left( t\right) $ is half the distance of the primaries,
\[
r\left( t\right) =\frac{1-e\cos E\left( t\right) }2.
\]
where the \emph{eccentric anomaly} $E\left( t\right) $ is the solution of the Kepler equation
\begin{equation}
E=t+e\sin E,  \label{kepler}
\end{equation}
and $e$ is the eccentricity of the ellipses described by the primaries. We will assume $0\le e\le 1$ all though
Subsections \ref{NonApprox} and \ref{Approx}. The Hamiltonian linked to the system is
\begin{equation}
H=\frac{v^2}2-\frac 1{\left( z^2\left( t\right) +r^2\left(
t\right) \right)^{1/2}},  \label{ghsitnikov}
\end{equation}
provided $v$ stands for $\dot{z}$ in the corresponding equations.

Since \eqref{ghsitnikov} cannot be solved in explicit form,
attempts at a Hamiltonian formulation of \eqref{sitnikov}, whether
exact or approximate, require one of at least two options: looking
for an exact Hamiltonian formulation by means of a change of
variables, which we will do in the next paragraph, and searching
an approximate Hamiltonian formulation, which will be done in
Subsection \ref{Approx}.

Let us now find a Hamiltonian linked to \ref{sitnikov}. We may express $r\left( t\right) $ as
\[
r\left( t\right) =\left( R\circ \varphi \right) \left( t\right) :=\frac{1-e^2}{2\left( 1+e\cos \varphi \left( t\right) \right) },
\]
where $\varphi $, the \emph{true anomaly}, is a solution of
\[
\frac{d\varphi }{dt}=\frac{\left( 1+e\cos \varphi \right) ^2}{\left(1-e^2\right) ^{3/2}}=\frac{\sqrt{1-e^2}}{4R^2\left( \varphi \right) },
\]
and we may follow the procedure introduced in \cite{Wodnar} (see also \cite{HagelTrenkler}) by taking $\varphi $
as the new independent variable and $x=\frac z{2r\left( \varphi \right) }$ as the new dependent variable. Writing
$t$ once again to denote $\varphi $, we have the following differential equation:
\begin{equation}
\ddot{x}=f\left( x,t\right) :=-\frac{e\cos t+\left( \frac 14+x^2\right)
^{-3/2}}{1+e\cos t}x,  \label{eqwodnar}
\end{equation}
clearly amenable to the hypotheses in Theorem \ref{th1} and in the first paragraph of Section \ref{mainresults}.
Defining $q_1=x,$ $q_2=t$ and $p_1=\dot{x}$, the autonomous Hamiltonian system corresponding to \eqref{eqwodnar}
is given by
\begin{equation}
\widehat{H}_e=H_e+p_2:=\frac{p_1^2}2+\frac{eq_1^2\cos q_2-4\left(
1+4q_1^2\right) ^{-1/2}}{2\left( 1+e\cos q_2\right) }+p_2,
\label{hsitnikov}
\end{equation}
always assuming $e\in \left[ 0,1\right] $.

The \emph{circular} Sitnikov problem $\widehat{H}_0$ is
meromorphically integrable in the sense of Liouville-Arnold and
can be solved using elliptic integrals. The non-integrability for
$e=1$ was first studied by means of straight Morales-Ramis theory
in \cite[\S 5]{moralesramisII} (see also \cite[\S 5.3]{Morales});
we will now extend the proof of meromorphic non-integrability
therein to one for every $0<e\le 1$ by using Theorem \ref{th1}.

The $\mathrm{NVE}_\Gamma $ derived from the Hamiltonian system given by \eqref{hsitnikov} is
\[
\ddot{\xi}=\left( \frac{e\left( 4q_1^2+1\right) ^{5/2}\cos q_2-64q_1^2+8}{
\left( e\cos q_2+1\right) \left( 4q_1^2+1\right) ^{5/2}}\right) \xi .
\]
Taking $q_1\equiv 0$ we have a solution
$\Gamma =\left\{ \mbox{\boldmath${z}$}\left( t\right)=\left( 0,t,0,\frac 2{1+e\cos t}\right) \right\} $ of
$X_{\widehat{H}_e}$ along which $\mathrm{NVE}_\Gamma $ is given by
\[
\ddot{\xi}=\left( \frac{e\cos t+8}{e\cos t+1}\right) \xi ,
\]
which is algebrizable, through the change of variable $\tau =\cos t$, into
\begin{equation}
\frac{d^2\xi }{d\tau ^2}-\left( \frac \tau {1-\tau ^2}\right) \frac{d\xi }{
d\tau }-\left( \frac{e\tau +8}{(e\tau +1)(1-\tau ^2)}\right) \xi =0.
\label{algesit3}
\end{equation}
This equation can be transformed into the differential equation
\begin{equation}
\frac{d^2\zeta }{d\tau ^2}=\left( \frac{5e\tau ^3+33\tau ^2-2e\tau -30}{
(e\tau +1)4(1-\tau ^2)^2}\right) \zeta ,  \label{algesit4}
\end{equation}
by means of $\xi =\frac \zeta {\sqrt[4]{1-\tau ^2}}$. Equations \eqref{algesit3} and \eqref{algesit4} are integrable
in terms of Liouvillian solutions only if $e=0$, whereas for $e\neq 0$ their solutions are given in terms of non-integrable
Heun functions if $e\neq 1$ and non-integrable Hypergeometric functions if $e=1$, i.e. their Galois groups are virtually
non-solvable, hence virtually non-abelian; furthermore, they have a regular singularity at infinity, which by Theorem
\ref{moralesramis} implies the non-integrability of $X_{\widehat{H}}$ for $e\in \left( 0,1\right] $ by means of
\emph{meromorphic} first integrals. In particular, the Galois group of equation \eqref{algesit4} is exactly
$\mathrm{SL}_2\left(\mathbb{C}\right) $.

\subsubsection{Numerical results for the Sitnikov Problem}

The author is indebted to Sergi Simon in what concerns the
following subsection, including the figures shown at the end of
the paper. Acknowledgments are also due to Carles Sim\'o for
further specific suggestions.

Let $\Sigma = \{  \sin\p{ q_2 } = 0 \}$. The six figures at the
end of this paper show Poincar\'e sections of the flow with
respect to $\Sigma$, projected on the $\p{q_1,p_1}$ plane, for the
Hamiltonian system $X_{H_e}$ obtained from \eqref{hsitnikov}. As
may be easily deduced from said Hamiltonian, all sections are
symmetrical with respect to the $q_1$ and $p_1$ axes. Different
amounts of initial conditions are used for the sake of clarity.

Figure \ref{fig1} corresponds to $e=0$. In keeping with what was
said after \eqref{hsitnikov}, the whole subset of $\Sigma$
transversal to the flow sheds concentric tori (ostensibly, the
intersections of the invariant Liouville-Arnold tori with
$\Sigma$), a typical sign of integrability; the tori shown are
only a selection of those therein, as the actual area foliated by
them is larger.

A number of these invariant tori break down upon the slightest
increase in $e$, and in the ensuing figures the two most
interesting features are those invariant sets (usually called
\emph{KAM tori}) whose intersection with $\Sigma$ prevails in the
form of Jordan curves, and the zones of chaotic behavior between
them. Sparse zones of the section will account for chaotic zones
as well, for $e>0$. Figures \ref{fig2} and \ref{fig3} show two
different close-up views for the Poincar\'e section corresponding
to $e=0.01$. The latter figure is actually a detail of the
``island'' of tori appearing at the right of the general section.
For $e= 0.1$, Figures \ref{fig4} and \ref{fig5} are, respectively,
a general view of the section and a close-up of one of the islands
appearing at each side of the central area. As for $e=0.4$, Figure
\ref{fig6} is an enlarged view of one of the two islands appearing
at each side of a central area.

\subsection{The approximate Sitnikov problem}\label{Approx}

As said in Subsection \ref{NonApprox}, we now consider an
approximation of the Sitnikov problem; see \cite{GideaDeppe} and
\cite{HeviaRanada} for more details. As opposed to
\emph{meromorphic} non-integrability, we will prove
non-integrability by means of \emph{rational} first integrals.

The fact that $\varphi \left( t\right) =t+O\left(e\right)$ yields
$r\left(t\right) =\frac{1-e\cos t}2+O\left( e^2\right)$, and thus
the Hamiltonian in \eqref{ghsitnikov} becomes
\[
H=\frac{v^2}2-\frac 1{\sqrt{z^2+\frac 14}}-e\frac{\cos t}{\left( z^2+\frac 14\right) ^{3/2}}+O\left( e^2\right)
\]
whenever $e\approx 0$. In particular, the first-order approximation of this asymptotic expansion in $e$ yields the
Hamiltonian
\begin{equation}
H=\frac{v^2}2-\frac 1{\sqrt{z^2+\frac 14}}-e\frac{\cos t}{\left( z^2+\frac 14\right) ^{3/2}}  \label{hsitnikovapprox}
\end{equation}

Considering $q_1=x,$ $q_2=t$ and $p_1=v$; the autonomous Hamiltonian system corresponding to this equation is given by
\begin{equation}
\widehat{H}=H+p_2,\quad H= {p^2_1 \over 2}-e{2\cos q_2\over(4 q_1^2 + 1)^{3/2}} - {2\over\sqrt{4 q_1^2 + 1}}, \label{hamth12}
\end{equation}
corresponding to the Hamiltonian in Theorem \ref{th1}:
$$f(x,t)=-{8x \over (4x^2+1)^{3 \over 2}}-e{24x\cos t \over (4x^2+1)^{5\over 2}}.$$
The $\mathrm{NVE}_{\Gamma}$ for the Hamiltonian \eqref{hamth12} is given by
$$\ddot \xi=\left(e{24 (16 q_1^2 - 1) \cos q_2\over (4 q_1^2 + 1)^{7\over 2}} + {8 (8 q_1^2 - 1)\over (4 q_1^2 + 1)^{5\over 2}}\right)\xi.$$
Its general solution may be expressed as: \[
\xi \left( t\right) =K_1C\left( 32,48e,\frac t2\right) +K_2S\left(
32,48e,\frac t2\right) ,
\]
where the \emph{Mathieu even} (resp. \emph{odd}) \emph{function} $C\left(
a,q,t\right) $ (resp. $S\left( a,q,t\right) $)\ is defined as the even
(resp. odd) solution to $\ddot{y}+\left( a-2q\cos \left( 2t\right) \right)
y=0$ (\cite[Ch. 20]{Abramowitz}).

Taking $q_1(t)=0$ we can see that $\bm{z}(t)=\left(0,t,0,2e\cos t+2\right)$;
hence, defining $\bm{z}(t)=\left(0,t,0,2e\cos t+2\right)$ and $\Gamma=\{\bm{z}(t)\}$,
the operator linked to $\mathrm{NVE}_{\Gamma}$
is given by
$$\ddot \xi=\left(- 24 e \cos t - 8\right)\xi,$$
which is algebrizable (see Theorem \ref{pr2}) through the change of variables $\tau=\cos t$ into
\begin{equation}\label{algesit1}
{d^2\xi\over d\tau^2}-\left({\tau\over 1- \tau^2}\right){d\xi\over d\tau}+\left({24e\tau+8\over 1-\tau^2}\right)\xi=0.
\end{equation}
Now, this equation can be transformed in the differential equation
\begin{equation}\label{algesit2}
{d^2\zeta\over d\tau^2}=\left({96 e\tau^3 + 31 \tau^2 - 96 e\tau - 34\over 4 (1-\tau^2)^2}\right)\zeta,\quad \xi={\zeta\over\sqrt[4]{1-\tau^2}}.
\end{equation}

Equations \eqref{algesit1} and \eqref{algesit2} are integrable in terms of Liouvillian solutions only if $e=0$,
since for $e\neq 0$ their solutions are given in terms of non-integrable Mathieu functions, hence their Galois
group are virtually non-solvable and thus virtually non-abelian; furthermore, they have an irregular singularity
at infinity, implying rational non-integrability for the Hamiltonian field $X_{\widehat{H}}$
with $\alpha=1$ in virtue of Theorem \ref{moralesramis}. In particular, the Galois group
of equation \eqref{algesit2} is exactly $\sl_2\p{\nc}$.

Equations \eqref{algesit1} and \eqref{algesit2} have been deeply analyzed in \cite{acbl} using the
Hamiltonian change of variables $\tau=e^{it}$, obtaining the same result presented here.

\subsection{Painlev\'e II equation: $\ddot x=2x^3+tx+\alpha$}
The non-integrability of the second Painlev\'e equation for
integer $\alpha$ has proved by Morales-Ruiz in \cite{mo1} and
later by Stoyanova et al in \cite{sto}. They used only the
Hamiltonian \eqref{hamth2}.

Defining $q_1=x,$ $q_2=t$, $p_1=y$ and $\alpha\in\mathbb{C}$´, the
autonomous Hamiltonian system corresponding to this equation can
given by any of the following three functions:
\begin{eqnarray}
\widehat{H}=H+p_2,\quad H= {\frac{p_1^2}2}-{q_1^4\over 2}-q_2{q_1^2\over 2}-\alpha q_1, \label{hamth1}\\
\widehat{H}=H+p_2,\quad H= {\frac{p_1^2}2}-\frac{1}{2}\left(q_1^2+{q_2\over 2}\right)^2-\alpha q_1, \label{hamcor2}\\
\widehat{H}=H+p_2,\quad H= {\frac{p_1^2}2}-\left(q_1^2+\frac{q_2}{2}\right)p_1-\left(\alpha+\frac{1}{2}\right)q_1, \label{hamth2}
\end{eqnarray}
where the equations \eqref{hamth1}, \eqref{hamcor2} and \eqref{hamth2} correspond to the Hamiltonian of
Theorem \ref{th1} ($f(x,t)=2x^3+tx+\alpha$), Corollary \ref{cor2} ($g(x)=x^2$ and $a(t)=t/2$) and Theorem \ref{th2}
($g(x)=x^2$ and $a(t)=t/2$) respectively. The Hamiltonian system for
Painlev\'e $\mathrm{II}$, studied in \cite{mo1,sto}, corresponds precisely to Hamiltonian \eqref{hamth2}. The
$\mathrm{NVE}_{\Gamma}$ for these Hamiltonians is given by $$\ddot \xi=\left(6q_1^2+q_2\right)\xi.$$

Taking $\alpha=0$ and $q_1(t)=0$ we have particular solutions $\bm{z}(t)=\left(0,t,0,0\right)$,
$\bm{z}(t)=\left(0,t,0,t^2/8\right)$ and $\bm{z}(t)=\left(0,t,t/2,t^2/8\right)$,
respectively, for the Hamiltonians \eqref{hamth1}, \eqref{hamcor2} and \eqref{hamth2}; hence $\mathrm{NVE}_{\Gamma}$
is given by $\ddot \xi=t\xi,$ the so-called \emph{Airy equation} (\cite[\S 10.4.1]{Abramowitz}),
which has an irregular singularity at infinity and is not integrable through Liouvillian solutions, i.e.
its Galois group is $\sl_2\p{\nc}$, not virtually abelian; thus, by Theorem \ref{moralesramis}, the Hamiltonian
field $X_{\widehat{H}}$ with $\alpha=0$ is not integrable through rational first integrals.

Now, for $\alpha=1$ and $q_1(t)=-1/t$, the integral curve $\bm{z}(t)$ is given by
$$\left(-{1\over t},t,{1\over t^2},-{1\over 2t}\right),\quad \left(-{1\over t},t,{1\over t^2},-{1\over 2t}+\frac{t^2}{8}\right) \textrm{  and  }\left(-{1\over t},t,{2\over t^2}+{t\over 2},-{1\over t}+{t^2\over 8}\right),$$
respectively for the Hamiltonians \eqref{hamth1}, \eqref{hamcor2} and \eqref{hamth2}, so that $\mathrm{NVE}_{\Gamma}$
is given by
$$\ddot \xi=\left({6\over t^2}+t\right)\xi,\quad \Gamma=\{\bm{z}(t)\},$$
whose general solution is
\[
\xi \left( t\right) =\sqrt{t}\left[ K_1I_{-5/3}\left( \frac{2t^{3/2}}
3\right) +K_2I_{5/3}\left( \frac{2t^{3/2}}3\right) \right] ,
\]
$I_{\alpha} =2^{-\alpha }t^{\alpha} \left( \frac 1{\Gamma \left( 1+\alpha
\right) }+\frac{t^2}{2^2\Gamma \left( 2+\alpha \right) }+O\left( t^4\right)
\right) $ being, for each $\alpha $, the \emph{modified Bessel function of
the first kind}, i.e. the solution to $t^2\ddot{y}+t\dot{y}-\left(
t^2+\alpha ^2\right) y=0$ (\cite[\S 9.6]{Abramowitz}).

The normal variational equation has an irregular singularity at infinity and is not integrable through
Liouvillian functions because its solutions are given in term of non-integrable Bessel functions (see \cite{mo1,sto}),
i.e. its Galois group is $\sl_2\p{\nc}$ which is not virtually abelian; again by the Morales-Ramis
Theorem \ref{moralesramis}, the Hamiltonian field $X_{\widehat{H}}$ with $\alpha=1$ is not integrable through
rational first integrals.

\subsection{An algebraic toy model: $\ddot x=-{1\over 4x^3}-{t\over x^2}+\alpha$}
Considering $q_1=x,$ $q_2=t$, $p_1=y$ and $\alpha\in\mathbb{C}$; the autonomous Hamiltonian systems corresponding to this equation are given by
\begin{eqnarray}
\widehat{H}=H+p_2,\quad H= {\frac{p_1^2}2}-{1\over 8q_1^2}-{q_2\over q_1}-\alpha q_1, \label{hamth11}\\
\widehat{H}=H+p_2,\quad H= {\frac{p_1^2}2}-\frac{1}{2}\left({1\over 2q_1}+2q_2\right)^2-\alpha q_1, \label{hamcor21}\\
\widehat{H}=H+p_2,\quad H= {\frac{p_1^2}2}+\left({1\over 2q_1}+2q_2\right)p_1-\left(\alpha+2\right)q_1. \label{hamth21}
\end{eqnarray}
\eqref{hamth11}, \eqref{hamcor21} and \eqref{hamth21} correspond to
Theorem \ref{th1} ($f(x,t)=-{1\over 4x^3}-{t\over x^2}+\alpha$),
Corollary \ref{cor2} ($g(x)=-{1\over 2x}$ and $a(t)=-2t$) and Theorem \ref{th2} ($g(x)=-{1\over 2x}$ and $a(t)=-2t$)
respectively. The $\mathrm{NVE}_{\Gamma}$ for all three is given by
$$\ddot \xi=\left({3\over 4q_1^4}+{2q_2\over q_1^3}\right)\xi.$$

Now, for $\alpha=1$ and $q_1(t)=\sqrt t$, the integral curve $\bm{z}(t)$ is given by
$$\left(\sqrt t,t,{1\over 2\sqrt t},2\sqrt t\right),\quad \left(\sqrt t,t,{1\over 2\sqrt t},2t^2+2\sqrt t\right) \textrm{  and  }\left(\sqrt t,t,-2t,2t^2\right),$$
respectively for the Hamiltonians \eqref{hamth11}, \eqref{hamcor21} and \eqref{hamth21}, rendering $\mathrm{NVE}_{\Gamma}$
equal to
\begin{equation}\label{nveex}
\ddot \xi=\left({3\over 4t^2}+{2\over \sqrt{t}}\right)\xi,,
\end{equation}
having a solution
\begin{eqnarray*}
\xi _1 &=&-\frac{3t^{3/2}}2\;_0F_1\left( ;\frac 73;\frac{8t^{3/2}}9\right)
\\
&=&-\frac 32t^{3/2}-\frac 47t^3-\frac 8{105}t^{9/2}+O\left( t^6\right),
\end{eqnarray*}
$_0F_1\left( ;a;t\right)=\lim_{q\rightarrow \infty} {}_1F_1\p{q;a;\frac{t}{q}}=\sum_{n= 0}^{\infty}\frac{t^n}{\p{a}_nn!}$
being the \emph{confluent hypergeometric limit function} (\cite[Ch. 13]{Abramowitz}),
and an independent new solution $\xi _2=\xi _1\int \xi _1^{-2}$, satisfying
\[
\xi _2=\frac 1{3\sqrt{t}}-\frac{8t}9-\frac{16}{27}t^{5/2}+O\left( t^4\right).
\]
As is the case for the rest of normal variational operators appearing in this paper, our knowledge of
the exponents around $0$ of a fundamental set of solutions (in this case, $\xi_1$ and $\xi_2$), coupled with the
basic result on factorization obtained in \cite[Th. 8 (Ch. 5)]{BoucherWeil}
(see also \cite[Criterion 1]{BoucherWeil}) would suffice
to prove non-integrability. Here, however, we will keep our restriction to
Theorems \ref{th1} and \ref{th2} and Corollary \ref{cor2}.

\eqref{nveex} is algebrizable (Theorem \ref{pr2}), through the change of variables $\tau=\sqrt t$ into
\begin{equation}\label{algej1}
{d^2\xi\over d\tau^2}-\left({1\over \tau}\right){d\xi\over d\tau}-\left({8\tau^3+3\over\tau^2}\right)\xi=0,
\end{equation}
now, this equation can be transformed in the differential equation
\begin{equation}\label{algej2}
{d^2\zeta\over d\tau^2}=\left({32\tau^3 + 15\over 4\tau^2}\right)\zeta,\quad \xi=\zeta\sqrt{\tau}.
\end{equation}

Equations \eqref{algej1} and \eqref{algej2} have an irregular singularity at $t=\infty$ and are not integrable
through Liouvillian solutions due to the presence of Bessel functions, i.e. their Galois group are virtually
non-solvable, therefore virtually non-abelian, Theorem \ref{moralesramis} once again settling
rational non-integrability for $\alpha=1$. In particular, the Galois group of equation \eqref{algej2} is exactly $\sl_2\p{\nc}$.

\section{Final Remarks: Open Questions and Future Work}
This paper is the starting point of a project in which the author
is involved. The following questions arose during our work:
\begin{itemize}
\item In \cite{mo1,sto} it was proven that the autonomous Hamiltonian
system related to Painlev\'e II is non-integrable for every
$\alpha\in\mathbb{Z}$. Is this also true for equation
(\ref{eq7})?
\item Does the integrability of equation (\ref{eq7}) for arbitrary $\alpha\in\mathbb{Z}$ depend on the choice of $g(x)$ and $a(t)$?
\item Assuming the above question has an affirmative answer, in what manner can the choice and form of $g(x)$ and $a(t)$
assure non-integrability for every $\alpha\in\mathbb{Z}$? and for every $\alpha\in\mathbb{C}$?
\item Is it possible to find transversal sections of the flow, and thus Poincar\'e maps, for either $\widehat{H}$
or the algebraized equation, even in the absence of non-trivial numerical monodromies? Do Stokes multipliers contribute to the answer in a significant manner?
\end{itemize}

Among our next goals, the analysis of the following items is due further immediate research:
\begin{itemize}
\item the application of Morales-Ramis theory to higher variational equations of non-autonomous Hamiltonian systems;
\item differential equations in the form $\ddot x=f(x,\dot x, t)$;
\item the rest of Painlev\'e equations: Casale in \cite{casale}
analyzed Painlev\'e I and Horozov et al in \cite{horozov} analized
some particular cases for Painlev\'e VI;
\item the theoretical aspects of non-autonomous Hamiltonian systems such as their geometry and the feasibility of an analogue to Liouville-Arnold theory;
\item the non-integrability of non-autonomous Hamiltonian systems with two and a half degrees of freedom;
\item specific examples of non-autonomous Hamiltonian systems related to control theory,
as well as other related to Celestial Mechanics, such as Restricted Three- and Four-Body Problems and
H\'enon-Heiles systems (\cite{Andreu}, \cite{GabernJorbaRobutel}).
\item the exact relation, perhaps causal, between separatrix splitting (\cite{MoralesSplitting}) and non-integrability, whether rational
or meromorphic.
\end{itemize}

\section*{Acknowledgments}
The research of the author is partially supported by grant FPI
Spanish Government, project BFM2003-09504-C02-02. The author is
indebted to Sergi Simon for multiple suggestions concerning
organization, style, pictures and numerical remarks, as well as
for preliminary corrections concerning the Sitnikov Problem; for
his collaboration in the numerical analysis thereof,
acknowledgments are also due to Carles Sim\'o. The author
acknowledges Juan J. Morales-Ruiz and Jacques-Arthur Weil for
their comments and suggestions and for the final proofreading of
this paper.
\newpage

\bibliographystyle{amsplain}
\nocite{*}
\def\cprime{$'$} \def\cs{Sim{\'o}, Carles}\def\mr{Morales-Ruiz, Juan
  Jos\'e}\def\cdprime{$''$}
\providecommand{\bysame}{\leavevmode\hbox
to3em{\hrulefill}\thinspace}
\providecommand{\MR}{\relax\ifhmode\unskip\space\fi MR }
\providecommand{\MRhref}[2]{%
  \href{http://www.ams.org/mathscinet-getitem?mr=#1}{#2}
} \providecommand{\href}[2]{#2}

\end{document}